\documentclass[aps,prd,amsmath,amssymb,showpacs]{revtex4}
\usepackage{amssymb}
\usepackage{mathbbol}
\usepackage{amsfonts}
\usepackage{mathrsfs}
\usepackage{epsfig,bm,dcolumn}
\usepackage{graphicx}
\usepackage{color}
\usepackage{amsmath}
\usepackage{dcolumn}
\usepackage{overpic}
\usepackage{hyperref}
\usepackage{slashed}

\usepackage[normalem]{ulem}

\newcommand{\stkout}[1]{\ifmmode\text{\sout{\ensuremath{#1}}}\else\sout{#1}\fi}

\begin{document}
\title{The quantum kinetic equation and dynamical mass generation in 2+1 Dimensions}
\author{Anping Huang$^{a,b}$}
\author{Shuzhe Shi$^c$}
\author{Xianglei Zhu$^{a}$}
\author{Lianyi He$^{d}$}
\author{Jinfeng Liao$^{b}$}\email{liaoji@indiana.edu}
\author{Pengfei Zhuang$^d$}\email{zhuangpf@mail.tsinghua.edu.cn}
\address{$^a$Department of Engineering Physics, Tsinghua University, Beijing 100084, China.\\
	$^b$Physics Department and Center for Exploration of Energy and Matter,	Indiana University, 2401 N Milo B. Sampson Lane, Bloomington, IN 47408, USA.\\
	$^c$Department of Physics, McGill University, Montreal, QC H3A 2T8, Canada.\\
	$^d$Department of Physics, Tsinghua University, Beijing 100084, China.\\
	}

\date{\today}

\begin{abstract}
In this work, we study the relativistic quantum kinetic equations in 2+1 dimensions from Wigner function formalism by carrying out a systematic semi-classical expansion up to $\hbar$ order.  The derived equations allow us to explore interesting transport phenomena in 2+1 dimensions. Within this framework, the parity-odd transport current induced by the external electromagnetic field is self-consistently derived. 
We also examine the dynamical mass generation by implementing four-fermion interaction with mean-field approximation. In this case, a new kind of transport current is found to be induced by the gradient of the mean-field condensate.  Finally, we also utilize this framework to study the dynamical mass generation in an external magnetic field for the  2+1 dimensional system under equilibrium. 
\end{abstract}

\maketitle

\section{Introduction}

The study of relativistic quantum kinetic theory has attracted much interest recently, partly motivated by the theoretical developments and experimental search for anomalous chiral transport phenomena in relativistic nuclear collisions~\cite{Bzdak:2019pkr, Kharzeev:2015znc,Kharzeev:2020jxw}.  Such a theory is both theoretically important as a many-body theoretical description and practically useful for describing relevant transport in a general out-of-equilibrium system. Many interesting results were obtained in 3+1 dimensions, such as the relativistic kinetic theory for scalar and fermions without external field~\cite{DeGroot:1980dk}, the anomalous chiral transport equation  in heavy-ion collisions~\cite{Son:2012zy,Gao:2012ix,Chen:2012ca,Hidaka:2016yjf,Huang:2018wdl,Wang:2019moi}, the quantum kinetic theory for massive fermions under external fields~\cite{Vasak:1987um,Zhuang:1995pd,Zhuang:1995jb,Zhuang:1998bqx,Chen:2013iga,Li:2019rth,Gao:2019znl,Weickgenannt:2019dks,Hattori:2019ahi,Dayi:2020uwx}, and the non-relativistic kinetic theory of spin-polarized system~\cite{Morawetz:2015iqd,Morawetz:2015fep,Zamanian:2010zz}. 

The relativistic quantum kinetic theory in 2+1 dimensions is of its own interest. Firstly, it could be a good starting point for developing a full quantum transport description for the case of massive fermions. At the moment,  it is still a challenge to derive the kinetic equation with the spin evolution of the massive fermions under the external Abelian gauge field in 3+1 dimensions. One technical reason is that there are 16 independent components of Wigner function in 3+1 D, which are coupled with each other by the mass term and very complicated to solve. In contrast,  there are just 4 independent components of Wigner function in 2+1 dimensions, and their equations are much simpler. This could allow a better conceptual and technical understanding of the finite mass effects in the quantum transport equations~\cite{Chen:2013dca}. Secondly, the Quantum ElectroDynamics in 2+1~D~(QED$_{3}$) has attracted recent physical interest, e.g. in the studies of the high-$T_{c}$ superconducting systems~\cite{Wen:1992ej,Rantner:2000wer} and the graphene~\cite{Khveshchenko:2001zz,Gusynin:2005pk,Lee:2007dzgxt,Feng:2006mwp}. Other interesting phenomena in 2+1 dimensions include e.g. the   fermion condensation in the massless limit induced by the magnetic fields $\langle\bar{\psi}\psi \rangle\varpropto |eB|$~\cite{Gusynin:1994re,Das:1995bn,Alexandre:2000yf,Raya:2010id}.  

In this work, we will study the relativistic quantum kinetic equations in 2+1 dimensions within the Wigner function formalism. The starting point is a Dirac theory with massive fermions coupled to external electromagnetic fields and with dynamical four-fermion interactions, i.e. the Nambu--Jona-Lasinio (NJL) model in 2+1~D which allows considering the dynamical mass generation in the external electromagnetic field. 
To systematically derive the kinetic equation of the NJL model in 2+1~D, we will adopt the strategy in our previous work~\cite{Huang:2018wdl}, starting from the Lagrangian of NJL model in 2+1~D and carrying out the semi-classical expansion by keeping the equations up to $\hbar$ order. In doing so,  we will self-consistently derive the well-known parity-odd transport current, $j^{\mu}\propto\,\epsilon^{\mu\rho\sigma}F_{\rho\sigma}$, \cite{Niemi:1983rq,Redlich:1983kn,Redlich:1983dv,Ishikawa:1983ad,Ishikawa:1984zv,Semenoff:1984dq,Bracken:2008zza}. Furthermore,
we will study the dynamical mass generation in this general framework by deriving and consistently solving the gap equation and the  kinetic equations together. The results will provide useful insights about the quantum effects beyond the mean-field as well as the role of the external magnetic field in dynamical mass generation.  As we shall show later, the massive fermions in 2+1 D demonstrate certain quantum features that would emerge in 3+1 D only for massless chiral fermions. The case here is similar to the study of Weyl fermions. For massless fermions, one usually study particles with parallel or anti-parallel spin and momentum in two dimensional spin space. It may be noted that in general, a physical fermion state in (2+1)D could be the superposition of two inequivalent irreducible representations as mirror images of each other. In the present work, we choose to  focus one sector  composed of the particles with ``spin-up'' and the anti-particles with ``spin-down'' (--- see Appendix A for further definitions).  Our study would help provide useful theoretical understanding for physical systems typically containing both sectors, and it might also be possible that certain future (2+1)D quantum materials might realize an isolated sector. 

The paper is organized as follows. In Sec.\ref{sec01}, we will give a simple review on Wigner function formalism and derive the full quantum kinetic equations in this approach without collision term. In Sec.\ref{sec02}, we focus on deriving the covariant transport equations in 2+1~D and the equal-time transport equations by carrying out the semi-classical expansion method, as well as self-consistently deriving the parity-odd transport currents.   In Sec. \ref{sec03}, we will further obtain the covariant and equal-time quantum transport equations with collision term in relaxation time approximation. In Sec.\ref{sec04}, we will investigate the dynamical mass generation under the external field from the gap equation that incorporates quantum effects.  Finally, we conclude in Se.\ref{sec05}.

\section{Equation of motion for the Wigner function}\label{sec01}

To study the dynamical mass generation for fermions in 2+1~Dimensions, let's consider the NJL model in 2+1~Dimensions, which can be written as the following form~\cite{Gomes:1989pp, Klimenko:1990rh, Klimenko:1992ch,Gusynin:1994re,Shovkovy:2012zn}
\begin{align}\label{eq:lagrangian1}
	\mathscr{L}=\bar{\psi}\left(i\hbar \gamma^\mu D_\mu - m_{0} \right)\psi+\frac{G}{2}(\bar{\psi}\psi)^2 \,,
\end{align}
where $D_{\mu}=\partial_{\mu}+iQA_{\mu}/\hbar$ is the covariant derivative, and the dimension of the charge $Q$ is $[m]^{1/2}$, which is different from the case in 3+1 D. Besides, there are some other differences between the 2+1 D and the 3+1 D, although the Lagrangian density takes the same formula. Firstly, the Dirac matrices $\gamma$ in 2+1 D are different from those in 3+1 D. There are two nonequivalent irreducible representations of the Dirac matrices in 2+1 D, which are characterized by $\frac{i}{2}\mathrm{Tr}(\gamma^{0}\gamma^{1}\gamma^{2})=s$, $s=\pm1$~\cite{Sitenko:1999jw}. In this work, we choose the Jackiw representation~\cite{Deser:1981wh}
\begin{align}\label{eq:gamma1}
	\gamma^{0}=\tau^{3},\quad
	\gamma^{1}=i\tau^{1},\quad
	\gamma^{2}=i\tau^{2},\quad
	\gamma^{\mu}\gamma^{\nu}=g^{\mu\nu}-i\epsilon^{\mu\nu\alpha}\gamma_{\alpha}.
\end{align}
Here, $\tau^{i}$ are the Pauli matrices and $\epsilon^{012}=\epsilon_{012}=1$, $g^{\mu\nu}=\mathrm{diag}(1,-1,-1)$. Particularly, $\{I_{2\times2}, \gamma^0, \gamma^1, \gamma^2\}$ form a complete, linearly independent basis of $2\times 2$ matrices, and the chirality $\gamma^{5}=i\gamma^{0}\gamma^{1}\gamma^{2}=-1$ has a fixed value in the Jackiw representation~\cite{Chen:2013dca}. The other nonequivalent irreducible representations is obtained by flipping the sign of $\gamma^{\mu}$, $\widetilde{\gamma}^{\mu}=-\gamma^{\mu}$, and $\widetilde\gamma^{5}=+1$. Secondly, unlike in 3+1 D, the spinor $\psi$ in Eq.(\ref{eq:lagrangian1}) just represents the particle with spin up or the anti-particle with spin down in the irreducible representations of Dirac matrices Eq.(\ref{eq:gamma1}), as discussed in \ref{appendix-spin}, while the spin-down particles and spin-up anti-particles are represented by the other irreducible representations characterized by $\widetilde{\gamma}^{\mu}$~\cite{deJesusAnguianoGalicia:2005ta}. Thirdly, as discussed in Ref.~\cite{Deser:1981wh}, the Lagrangian density in Eq.(\ref{eq:lagrangian1}) is not invariant under parity transformation, due to the mass term with the irreducible representations Eq.(\ref{eq:gamma1}). To see this, we take the parity transformation corresponding to flipping the sign of one of axes, say $\hat{x}$, and a spinor under such parity transformation is $\widehat{\mathrm{P}}_x \psi(t,x,y)\widehat{\mathrm{P}}_x=-i\gamma^{1}\psi(t,-x,y)$, then the mass term will flip the sign under parity transformation, i.e $\widehat{\mathrm{P}}_x\bar{\psi}\psi(t,x,y)\widehat{\mathrm{P}}_x = -\bar{\psi}\psi(t,-x,y)$.

In this work, we choose one of the irreducible representations of Dirac matrices, which means we focus on the sub-system composed of the particles with spin up and the anti-particle with spin down. Consequently, the properties of positive charge and negative charge are not necessarily the same. For instance, the dynamical mass is different for opposite charges. Besides, this system is similar to the chiral system, the spin is locked in both of them. In the chiral system, the spin is either parallel or anti-parallel to the momentum, while the spin is out-of-plane in this system.

Under the mean field approximation, the Lagrangian density can be reduced as  
\begin{align}
	\mathscr{L}_{MF}=\bar{\psi}\left(i\hbar \gamma^\mu D_\mu -(m_{0}+\sigma) \right)\psi-\frac{1}{2G}\sigma^2 \,,
\end{align}
with the effective mass $\sigma=-G\langle \bar{\psi}\psi \rangle$. In this work, the flavor structure is not considered for simplification. From the mean field effective Lagrangian density, one obtains the Dirac equation:
\begin{align*}
	\left[i\hbar\gamma^\mu D_\mu - (m_{0}+\sigma) \right]\psi=0,\qquad
	\bar{\psi}\left[i\hbar\gamma^{\mu}D^{+}_{\mu} + (m_{0}+\sigma) \right]=0.
\end{align*}
Where the operator $D^{+}_{\mu} = \overleftarrow{\partial}_{\mu}-iQA_{\mu}/\hbar$, $\overleftarrow{\partial}_{\mu}$ is the space-time derivative which acts only on the former function.

The covariant and gauge invariant Wigner function for fermions in 2+1 dimensions is~\cite{Vasak:1987um,Chen:2013dca} 
\begin{align}
	W_{\alpha\beta}(x,p)=\int\frac{d^{3}y}{(2\pi\hbar)^{3}} e^{-\frac{i}{\hbar}p\cdot y} \left< \bar{\psi}_{\beta}(x_{+})U(x,y)\psi_{\alpha}(x_{-})\right>,
	\qquad U(x,y)= e^{-\frac{i}{\hbar}Q \int^{x_{+}}_{x_{-}}dz^{\mu}A_{z}},
\end{align}
where the notation $x_{\pm}=x\pm \frac{y}{2}$, and the function $U(x,y)$ is the gauge link ensuring the invariance under gauge transformation. 
Combining the above Dirac equations and the definition of Wigner function, we can write down the equation of motion for Wigner function as~\cite{Florkowski:2018ahw,Zhuang:1995jb},
\begin{align}\label{eq:EOMW}
	\left(\slashed{K}-M\right)W(x,p)=0 \,,
\end{align}
where the operator $K^{\mu}=\pi^{\mu}+\frac{1}{2}i\hbar\triangledown^{\mu}$, herein these two operators are respectively,
\begin{align*}
	\pi^{\mu}=p^{\mu}-\frac{1}{2}Q\hbar\,j_{1}\left(\frac{1}{2}\hbar\triangle\right)F^{\mu\nu}\partial^{p}_{\nu},\qquad
	\triangledown^{\mu}=\partial^{\mu}-Q\,j_{0}\left(\frac{1}{2}\hbar\triangle\right)F^{\mu\nu}\partial^{p}_{\nu},
\end{align*} 
and the function $j_{i} (i=0,1)$ is the spherical Bessel function. The mass $M$ operator can be decomposed as the 
\begin{align}
	M=M_{1}-iM_{2},\qquad
	M_{1}=m_{0}+\cos\left(\frac{1}{2}\hbar\triangle\right)\sigma,\qquad
	M_{2}=\sin\left(\frac{1}{2}\hbar\triangle\right)\sigma.
\end{align} 
Here the triangle operator $\triangle=\partial_x\cdot\partial_p$, in which the spatial derivative $\partial_{x}$ only acts on the effective mass $\sigma$, but not on the Wigner function $W$.

The Wigner function is a $2\times 2$ matrix, and it can be expanded in terms of 4 independent generators of Clifford algebra,
\begin{align}\begin{split}\label{eq:wc}
		&W(x,p)=\frac{1}{2}\left(\mathscr{F}+\gamma^{\mu}V_{\mu} \right),\\
		&\mathscr{F}=\mathrm{Tr}[W],\qquad \mathscr{V}^{\mu}=\mathrm{Tr}[\gamma^{\mu}W].
\end{split}\end{align}
It is obviously that the Wigner function in 2+1~D has less independent degrees of freedom compared to that in 3+1~D --- the pseudo-scalar, axial-vector and antisymmetric tensor are absent in 2+1~D.  These four coefficients correspond to some physical distributions---the mass density, current density and energy-momentum tensor density: 
\begin{align}\label{eq:dce}
	\sigma=-G\int d^{3}p\mathscr{F}(x,p),\qquad
	j^{\mu}(x)=\int \mathrm{d}^{3}p\,\mathscr{V}^{\mu}(x,p),\qquad
	T^{\mu\nu}(x)=\int \mathrm{d}^{3}p\,p^{\mu}\mathscr{V}^{\nu}(x,p).
\end{align}

According to Noether's theorem, the conserved angular-momentum flux density is
\begin{align*}
&J^{\lambda\mu\nu}=x^{\mu}T^{\lambda\nu}-x^{\nu}T^{\lambda\mu}+S^{\lambda\mu\nu}.
\end{align*}
Herein, the first two terms represent the orbital part of the angular momentum, which depend on the canonical energy-momentum tensor density $T^{\mu\nu}$. While the last term defines the canonical spin tensor density, which can be written as as~\cite{Florkowski:2018ahw,Binegar:1981gv,Belich:2001hf}
\begin{align}
    S^{\mu\alpha\beta} \equiv \frac{\hbar}{4} \left< \bar{\psi}(x)\{\gamma^{\mu},\sigma^{\alpha\beta}\} \psi(x)\right>
    =\frac{\hbar}{4}\int d^{3}p\,\mathrm{Tr}\left[\{\gamma^{\mu},\sigma^{\alpha\beta}\}W(x,p)\right]
    =\frac{\hbar}{2}\epsilon^{\mu\alpha\beta}\int \mathrm{d}^{3}p \mathscr{F}(x,p),
\end{align}
where, the spin information is encoded in the scalar component $\mathscr{F}(x,p)$.
It is more clearly by the following relation,
\begin{align*}
S^{0ij}=\frac{\hbar}{4}\langle\bar{\psi}\{\gamma^{0}, \sigma^{ij}\}\psi\rangle
=\hbar\,\epsilon^{0ij}\langle\psi^{\dagger}\frac{\sigma_{z}}{2}\psi\rangle
=\frac{\hbar}{2}\epsilon^{0ij} \int d^{3}p\, \mathscr{F}(x,p).
\end{align*}

Substituting Eq.~(\ref{eq:wc}) into Eq.~(\ref{eq:EOMW}), one can derive the kinetic equations for the four independent components as follows.
\begin{align}\begin{split}\label{eq:wcq}
		&\pi^{\mu}\mathscr{V}_{\mu}-M_{1}\mathscr{F}=0,\\
		&\frac{1}{2}\hbar\triangledown^{\mu}\mathscr{V}_{\mu}+M_{2}\mathscr{F}=0,\\
		&\pi_{\mu}\mathscr{F}-M_{1}\mathscr{V}_{\mu}+\frac{1}{2}\hbar\epsilon_{\mu\rho\sigma}\triangledown^{\rho}\mathscr{V}^{\sigma}=0,\\
		&\frac{1}{2}\hbar\triangledown_{\mu}\mathscr{F}+M_{2}\mathscr{V}_{\mu}-\epsilon_{\mu\rho\sigma}\pi^{\rho}\mathscr{V}^{\sigma}=0.
\end{split}\end{align} 
These equations form the complete equation set to describe the evolution of the system. Although they are much simpler than those in 3+1~D, these equations are still hard to solve. Similar to our previous work~\cite{Huang:2018wdl}, for the rest of this paper, we will take the semi-classical approximation and expand the equations in orders of $\hbar$ to simplify the above equations.

\section{Transport Equation without collision term}\label{sec02}
Our goal is to derive the quantum kinetic equation of the fermions under external Abelian field from Eq.~(\ref{eq:wcq}). As did in~\cite{Huang:2018wdl}, we take the semi-classical expansion to the operators and the components of Wigner function as following 
\begin{align}
&\pi^{\mu}=p^{\mu}-\frac{1}{12}Q\hbar^{2}\triangle\,F^{\mu\nu}\partial^{p}_{\nu}+\mathcal{O}(\hbar^{4}),~~\triangledown^{\mu}=\partial^{\mu}-QF^{\mu\nu}\partial^{p}_{\nu}+\mathcal{O}(\hbar^{2}),\\
&M_{1}=m-\frac{1}{2}\left(\frac{1}{2}\hbar\triangle\right)^{2}\sigma+\mathcal{O}(\hbar^{4}),~~~~M_{2}=\frac{1}{2}\hbar\triangle\sigma+\mathcal{O}(\hbar^{3}),\\
&\mathscr{F}=\mathscr{F}^{0}+\hbar\mathscr{F}^{1}+\mathcal{O}(\hbar^{2}),~~\mathscr{V}^{\mu}=\mathscr{V}^{\mu}_{0}+\hbar\mathscr{V}^{\mu}_{1}+\mathcal{O}(\hbar^{2}).
\end{align}
In above equations, $m=m_{0}+\sigma$, is the effective mass, while $\sigma=-G\int d^{3}p\mathscr{F}(x,p)$. Now, we can solve the Eq.~(\ref{eq:wcq}) order by order.

\subsection{The zeroth order}
To the zeroth order, Eq.~(\ref{eq:wcq}) can be written as
\begin{align}\begin{split}\label{eq:wcq0}
&p^{\mu}\mathscr{V}_{\mu}^{(0)}-m\mathscr{F}^{(0)}=0,\\
&\triangledown^{\mu}\mathscr{V}^{(0)}_{\mu}+\triangle\sigma(x)\mathscr{F}^{(0)}=0,\\
&p_{\mu}\mathscr{F}^{(0)}-m\mathscr{V}^{(0)}_{\mu}=0,\\
&\epsilon_{\mu\rho\sigma}p^{\rho}\mathscr{V}^{\sigma}_{(0)}=0.
\end{split}\end{align} 
From the first and third equations of the above set of equations, one can get the on-shell condition for $\mathscr{F}$,
\begin{align}
	&( p^{2}-m^{2}) \mathscr{F}^{(0)}=0.
\end{align}
We can formally write the $\mathscr{F}^{(0)}$ as 
\begin{align}\label{eq:F01}
	&\mathscr{F}^{(0)}=m\,f^{(0)}(x,p)\delta(p^{2}-m^{2}).
\end{align}
Then the vector $\mathscr{V}^{\mu}_{(0)}$ can be represented as
\begin{align}\label{eq:v0}
	&\mathscr{V}^{\mu}_{(0)}=p^{\mu}f^{(0)}(x,p)\delta(p^{2}-m^{2}).
\end{align}
According to the definition of the current density Eq.~(\ref{eq:dce}), we can get 
\begin{align}
	&j^{\mu}_{0}=\int d^{3}p^{\mu}f^{(0)}(x,p)\delta(p^{2}-m^{2}).
\end{align}
Then the physical meaning of function $f^{(0)}(x,p)$ now is clear--- it can be interpreted as the zeroth order  distribution function of the fermions in 2+1~D.  


It might be worth noting that  the equation set~(\ref{eq:wcq0}) contains four equations: we obtain the formal solution of $\mathscr{F}^{(0)}$ and $\mathscr{V}^{\mu}_{(0)}$ by using the first and third equations, then the fourth equation is automatically satisfied. Meanwhile, the second equation of (\ref{eq:wcq0}) leads to the evolution equation of $f^{(0)}(x,p)$, i.e. the zeroth order covariant transport equation: 
\begin{align}\label{eq:tq0}
&\delta(p^{2}-m^{2})\left(p\cdot \triangledown+m\sigma_{\nu}\partial^{\nu}_{p}\right)f^{(0)}(x,p)=0.
\end{align}
Where the operator $\triangledown^{\mu}=\partial^{\mu}-QF^{\mu\nu}\partial^{p}_{\nu}$, and herein we have introduced a notation $\sigma_{\nu}=\partial_{\nu}\sigma(x)$. The corresponding Gap equation can be written as 
\begin{align}\label{eq:gap0}
	&m-m_{0}=-G\,m\int \mathrm{d}^{3}p\,f^{(0)}(x,p)\delta(p^{2}-m^{2}).
\end{align}

These above two equations Eq.(\ref{eq:tq0},\ref{eq:gap0}) form a complete, self-consistent kinetic transport equation at zeroth order. They should be solved concurrently when solving the transport equations numerically. Now the information of zeroth order is clear. With this, we move on to construct the kinetic equation up to the order of $\hbar$.

\subsection{The first order}
The $\hbar$-order sector of Eq.~(\ref{eq:wcq}) is 
\begin{align}\begin{split}\label{eq:wcq1}
&p^{\mu}\mathscr{V}_{\mu}^{(1)}-m\mathscr{F}_{(1)}=0,\\
&\triangledown^{\mu}\mathscr{V}^{(1)}_{\mu}+\triangle\sigma(x)\mathscr{F}^{(1)}=0,\\
&p_{\mu}\mathscr{F}^{(1)}-m\mathscr{V}^{(1)}_{\mu}+\frac{1}{2}\epsilon_{\mu\rho\sigma}\triangledown^{\rho}\mathscr{V}^{\sigma}_{(0)}=0,\\
&\frac{1}{2}\triangledown_{\mu}\mathscr{F}_{(0)}+\frac{1}{2}\triangle\sigma(x)\mathscr{V}^{(0)}_{\mu}-\epsilon_{\mu\rho\sigma}p^{\rho}\mathscr{V}^{\sigma}_{(1)}=0,
\end{split}\end{align} 
where the operator $\triangledown^{\mu}=\partial^{\mu}-QF^{\mu\nu}\partial^{p}_{\nu}$. According to the first and third equation in Eq.~(\ref{eq:wcq1}), one can get 
\begin{align}\label{eq:F11}
&(p^{2}-m^{2})\mathscr{F}^{(1)}+\frac{1}{2}\epsilon^{\mu\rho\sigma}p_{\mu}\triangledown_{\rho}\mathscr{V}^{(0)}_{\sigma}=0.
\end{align}
Plugging in the solution of $\mathscr{V}^{(0)}_{\sigma}$~(\ref{eq:v0}), the second term of above equation is
\begin{align*}
&\frac{1}{2}\epsilon^{\mu\rho\sigma}p_{\mu}\triangledown_{\rho}\mathscr{V}^{(0)}_{\sigma}=-Qp\cdot \widetilde{F}\,f^{(0)}(x,p)\delta(p^{2}-m^{2}).
\end{align*}
Herein, $\widetilde{F}^{\mu}=\frac{1}{2}\epsilon^{\mu\rho\sigma}F_{\rho\sigma}=(-B,-E^{2},E^{1})$ is the dual field strength, and $B=-1/2\epsilon^{ij}F_{ij}$, $F_{ij}=-\epsilon_{ij}B$, $E^{i}=F^{i0}$. It is interesting that the magnetic field $B$ is a pseudo-scalar rather than a pseudo-vector.

Now Eq.~(\ref{eq:F11}) can be further simplified as
\begin{align}
	&(p^{2}-m^{2})\mathscr{F}^{(1)}=Qp\cdot \widetilde{F}\,f^{(0)}(x,p)\delta(p^{2}-m^{2}).
\end{align}
Utilizing the property of delta function, $x\delta^{'}(x)=-\delta(x)$, the solution of $\mathscr{F}^{(1)}$ can be formally written as 
\begin{align}
	&\mathscr{F}^{(1)}=G(x,p)\delta(p^{2}-m^{2})-Qp\cdot \widetilde{F}\,f^{(0)}(x,p)\delta^{'}(p^{2}-m^{2}).
\end{align}
In addition, we have introduced a new function $G(x,p)$. From the third equation in Eq.~(\ref{eq:wcq1}), we can get the solution of $\mathscr{V}^{\mu}_{1}$,
\begin{align}\label{eq:v11}
	\mathscr{V}^{\mu}_{(1)}&=\frac{1}{m}p^{\mu}G(x,p)\delta(p^{2}-m^{2})-\frac{1}{m}Q\,p^{\mu}p\cdot \widetilde{F}\,f^{(0)}(x,p)\delta^{'}(p^{2}-m^{2})+\frac{1}{2m}\epsilon^{\mu\rho\sigma}\triangledown_{\rho}V^{(0)}_{\sigma}.
\end{align}
After some calculation, the last term can be reduced as 
\begin{align*}
	&\frac{1}{2m}\epsilon^{\mu\rho\sigma}\triangledown_{\rho}V^{(0)}_{\sigma}\\
	&=\frac{1}{2m}\epsilon^{\mu\rho\sigma}p_{\sigma}\left(\triangledown_{\rho} f^{(0)}(x,p)\right)\delta(p^{2}-m^{2})+\frac{Q}{m}p^{\mu}p\cdot \widetilde{F}\,f^{(0)}(x,p)\delta^{'}(p^{2}-m^{2})-mQ\widetilde{F}^{\mu}f^{(0)}\delta^{'}(p^{2}-m^{2})\\
	&~~-\epsilon^{\mu\rho\sigma}p_{\sigma}\sigma_{\rho}\,f^{(0)}(x,p)\delta^{'}(p^{2}-m^2).
\end{align*}
Here, the Schouten identity in 2+1~D,
\begin{align}\label{eq:Schouten}
	&p^{\lambda}\epsilon^{\mu\rho\sigma}-p^{\mu}\epsilon^{\rho\sigma\lambda}+p^{\rho}\epsilon^{\sigma\lambda\mu}-p^{\sigma}\epsilon^{\lambda\mu\rho}=0,
\end{align}
is employed.


Now, Eq.~(\ref{eq:v11}) can be further reduced as
\begin{align}\begin{split}
\mathscr{V}^{\mu}_{(1)}&=\frac{1}{m}p^{\mu}G(x,p)\delta(p^{2}-m^{2})+\frac{1}{2m}\epsilon^{\mu\rho\sigma}p_{\sigma}\left(\triangledown_{\rho} f^{(0)}(x,p)\right)\delta(p^{2}-m^{2})-mQ\widetilde{F}^{\mu}f^{(0)}(x,p)\delta^{'}(p^{2}-m^{2})\\
&~~-\epsilon^{\mu\rho\sigma}p_{\sigma}\sigma_{\rho}\,f^{(0)}(x,p)\delta^{'}(p^{2}-m^2)\,,
\end{split}\end{align}
and the physical meaning of the function $G(x,p)$ becomes clear. According to the definition of current density, the function $G(x,p)$ can be regarded as first order correction to the distribution function, and it can be formally decomposed as
\begin{align}
	&G(x,p) = m\,f^{(1)}(x,p).
\end{align}
Then the solution of $\mathscr{F}^{(1)} $ and $V^{\mu}_{(1)}$ can now be rewritten as

\begin{align}\begin{split}\label{eq:solution-1}
\mathscr{F}^{(1)}&=m\,f^{(1)}(x,p)\delta(p^{2}-m^{2})-Qp\cdot \widetilde{F}\,f^{(0)}(x,p)\delta^{'}(p^{2}-m^{2}),\\
\mathscr{V}^{\mu}_{(1)}&=p^{\mu}f^{(1)}(x,p)\delta(p^{2}-m^{2})+\frac{1}{2m}\epsilon^{\mu\rho\sigma}p_{\sigma}\left(\triangledown_{\rho} f^{(0)}(x,p)\right)\delta(p^{2}-m^{2})-mQ\widetilde{F}^{\mu}f^{(0)}(x,p)\delta^{'}(p^{2}-m^{2})\\
&~~-\epsilon^{\mu\rho\sigma}p_{\sigma}\sigma_{\rho}\,f^{(0)}(x,p)\delta^{'}(p^{2}-m^2).
\end{split}\end{align}

Using the formal solution of $\mathscr{V}^{\mu}_{(1)}$ , we can get the first order current density 
\begin{align}\begin{split}\label{eq:cur1}
j^{\mu}_{{1}}&=\int d^{3}p\mathscr{V}^{\mu}_{(1)}\\
&=\int d^{3}p\,p^{\mu}f^{(1)}(x,p)\delta(p^{2}-m^{2})+\frac{1}{2m}\epsilon^{\mu\rho\sigma}\int d^{3}p\,p_{\sigma}\left(\triangledown_{\rho} f^{(0)}(x,p)\right)\delta(p^{2}-m^{2})\\
&~~-mQ\widetilde{F}^{\mu}\int d^{3}p\,f^{(0)}(x,p)\delta^{'}(p^{2}-m^{2})-\epsilon^{\mu\rho\sigma}\sigma_{\rho}\int d^{3}p\,p_{\sigma}\,f^{(0)}(x,p)\delta^{'}(p^{2}-m^2).
\end{split}\end{align}
A couple of interesting physical phenomena can be seen in this equation. 
Firstly, let us focus on the third term, which is originated from the external electromagnetic field. In this term, the vacuum contributes a conserved vector current~\cite{Niemi:1983rq,Semenoff:1984dq,Redlich:1983kn,Redlich:1983dv,Ishikawa:1983ad,Ishikawa:1984zv}, 
\begin{align}\label{eq:curvac}
	&j^{\mu}_{v}=\hbar\,mQ\widetilde{F}^{\mu}\frac{2}{(2\pi)^2}\int d^{3}p\theta(-p_{0})\delta^{'}(p^{2}-m^{2})=\hbar\frac{Q}{8\pi}\epsilon^{\mu\alpha\beta}F_{\alpha\beta}.
\end{align} 
However, it is different form the normal electric current. Writing down different components explicitly, $j^{0}_{v}\propto\,B$, $j^{1}_{v}\propto\,E^{2}$, $j^{2}_{v}\propto\,E^{1}$, we find the current is perpendicular to the electric field. Also, it explicitly violates parity symmetry since $\widetilde{F}^{\mu}\equiv\epsilon^{\mu\alpha\beta}F_{\alpha\beta}$ is an axial-vector. 

Secondly, the last term in Eq.~(\ref{eq:cur1}) is a novel current, which is induced by the space-time gradient of the condensation, 
\begin{align}\begin{split}
&j^{\mu}_{\sigma}=-\hbar\epsilon^{\mu\rho\sigma}\sigma_{\rho}\int d^{3}p\,p_{\sigma}\,f^{(0)}(x,p)\delta^{'}(p^{2}-m^2)=-\hbar\,\epsilon^{\mu\rho\nu}\sigma_{\rho} u_\nu I_{m},\\
&I_{m}=\int d^{3}p\,(u \cdot p)\,f^{(0)}(x,p)\delta^{'}(p^{2}-m^2)
=-\sum_{\epsilon=\pm1}\epsilon\int\,\frac{d^{2}p}{(2\pi)^{2}2E^{(0)}_{p}}\frac{d}{dE^{(0)}_{p}}f_\mathrm{fluid}^{(0)\epsilon}(x,\mathbf{p}),
\end{split}\end{align}
where $u_\nu$ is the fluid velocity, and $f_\mathrm{fluid}$ is the distribution function in the fluid co-moving frame.
Unlike Eq.~(\ref{eq:curvac}), there is no vacuum contribution here, but only the medium contribution. However, it is similar to the current in Eq.~(\ref{eq:curvac}) in terms of its direction. Taking the rest frame of fluid, i.e. $u^\mu=(1,0,0)$, one can find $j^{0}_{\sigma}\propto\,0$, $j^{1}_{\sigma}\propto\,\sigma_{2}$, and $j^{2}_{\sigma}\propto\,\sigma_{1}$. Such current is perpendicular to the gradient of condensation function $\sigma(x)$.
It is clear that this gradient current vanishes when the condensation is homogeneous or when there is not net particle number. This new gradient current may be the special case in 2+1 D. It is a natural and interesting question to ask whether and how these nontrivial currents would also emerge in a fluid dynamic description of the same 2+1D massive fermion systems. Fluid dynamics with anomalous currents is known in the case of 3+1D chiral fermion systems \cite{Son:2009tf} and has important phenomenological applications \cite{Shi:2017cpu,Shi:2019wzi}. It will be tempting to construct a fluid dynamics for 2+1D massive fermions in a future study.

Next, we move on to discuss the transport equation for $f^{(1)}(x,p)$, which is determined by the second equation of Eq.~(\ref{eq:wcq1}). Substituting the solution for $\mathscr{V}^{\mu}_{(1)}$ into the second equation of Eq.~(\ref{eq:wcq1}, and applying appropriate simplifications, we find the equation of motion
\begin{align}\label{eq:tq12}
	&\delta(p^{2}-m^{2})\left[p\cdot\triangledown\,f^{(1)}(x,p)+\frac{1}{2m}Q(\partial_{\nu}p\cdot\widetilde{F})\left(\partial^{\nu}_{p}f^{(0)}(x,p) \right)-\frac{1}{2m^{2}}\epsilon^{\mu\rho\sigma}\sigma_{\mu}p_{\sigma}\left(\triangledown_{\rho}f^{(0)}(x,p)\right)+m\,\sigma\cdot\partial_{p}f^{(1)}(x,p)\right]\nonumber\\
	&-\delta^{'}(p^{2}-m^{2})\left[\frac{Q}{m}p\cdot \widetilde{F}\,\left(p\cdot\triangledown\,f^{(0)}(x,p)\right)+Q\,p\cdot\widetilde{F}\,\sigma\cdot\partial_{p}f^{(0)}(x,p)\right]=0.
\end{align}

\subsection{Covariant transport equation up to $\hbar$ order}


Now, let's combine the zeroth order and the first order transport equations, i.e Eq.~(\ref{eq:tq0}) and Eq.~(\ref{eq:tq12}), as well as the gap equation, we can get the complete covariant transport equation for fermions in 2+1~D as follows:
\begin{align}\begin{split}\label{eq:tq}
&\delta\left(p^{2}-m^{2}-\hbar\frac{Q}{m}p\cdot \widetilde{F}\right)\left[p\cdot\triangledown+m\sigma_{\nu}\partial^{\nu}_{p}+\hbar\frac{Q}{2m}\left(\partial_{\nu}p\cdot\widetilde{F}\right)\partial^{\nu}_{p}-\hbar\frac{1}{2m^{2}}\epsilon^{\mu\rho\sigma}\sigma_{\mu}p_{\sigma}\triangledown_{\rho}\right]f(x,p)=0 \,,\\
&m-m_{0}=-G\left[m\int \mathrm{d}^{3}p\,f(x,p)\delta(p^{2}-m^{2})-\hbar\,Q\int \mathrm{d}^{3}p\,p\cdot\widetilde{F}\,f^{(0)}(x,p)\delta^{'}(p^{2}-m^{2})\right] \,,
\end{split}\end{align}
where the distribution function $f(x,p)=f^{(0)}(x,p)+\hbar\,f^{(1)}(x,p)$, and we have used the Taylor expansion to the delta function in the transport equation with keeping to the first order. It is worth noting that the on-shell condition has been modified by the quantum effect. The correction is originated from the coupling between fermion's magnetic moment and the external electromagnetic field. According to the modified on-shell condition, we can get the shifted energy
\begin{align}
&p_{0}=\epsilon\,E_{p},~~E_{p}=E^{(0)}_{p}+\hbar\frac{\epsilon\,Q}{2m}\frac{\tilde{p}\cdot\widetilde{F}}{E^{(0)}_{p}}.
\end{align}
Herein, $\tilde{p}^{\mu}=(E^{(0)}_p,\epsilon\,\vec{p})$, and $E^{(0)}_{p}=\sqrt{\mathbf{p}^{2}+m^{2}}$ is the classical energy; $\epsilon=\pm1$ denotes the positive and negative energy respectively, which also means that the distribution now can be decomposed of two branches as follows,
\begin{align}\begin{split}\label{eq:fc}
	&f(x,p)=\frac{2}{(2\pi)^2}\sum_{\epsilon=\pm1}\theta(\epsilon\,p_{0})\widetilde{f}^{\epsilon}(x,\epsilon\,p),\\
	&\widetilde{f}^{+}(x,p)=f^{+}(x,p),~~\widetilde{f}^{-}(x,-p)=f^{-}(x,-p)-1.
\end{split}\end{align}
In these equations, we have included the vacuum contribution, because it contributes to the physics we are interested in, and $f^{\epsilon}(x,\epsilon\,p)$ is the particle ($\epsilon=1$) or anti-particle ($\epsilon=-1$) distribution function.

The quantum correction in energy is caused by the interaction between the fermion's magnetic moment and the external field. This is more clear in the particle co-moving frame, in which $\widetilde{p}^{\mu}=(p_0,0,0)$, and the energy becomes
\begin{align}
&E_{p}=E^{(0)}_{p}-\hbar\frac{\epsilon\,Q}{2m}B=E^{(0)}_{p}-\hbar\,\mu_{B}\,B.
\end{align}
$\mu_{B}=\frac{\epsilon\,Q}{2m}$ is the Bohr magneton. This is the Zeeman effect in 2+1~Dimension. It is interesting that the above shifted energy can also be treated as modification of the effective mass
\begin{align}
&E_{p}=\sqrt{\mathbf{p}^{2}+M^{2}},~~~M=m+\delta\,m,~~~\delta\,m=\hbar\frac{\epsilon\,Q}{2m^2}p\cdot\widetilde{F} \xrightarrow{\mathbf{E}=0}-\hbar\mu_{B}\frac{E^{(0)}_{p}}{m}B,
\end{align}
where M is the effective mass of the fermions under external field, while $\delta\,m$ is the mass correction which is at $\hbar$ order, and proportional to the magnetic field in the absence of  electric field $\mathbf{E}$.

\subsection{Equal-time transport equation}

In practical calculations of solving the transport equation numerically, we need the equal-time transport equation. The equal-time transport equation can be obtained by integration over $p_{0}$ to covariant transport equation in Eq.~(\ref{eq:tq}). After integration over $p_{0}$ and using the chain rule to the space-time and momentum derivatives (because of the energy $E_p$ no longer an independent variable), as well as replacing $\mathbf{p}$ by $\epsilon\mathbf{p}$, we can get 
\begin{align}\begin{split}
&\sum_{\epsilon=\pm1}\frac{\epsilon}{2}E_{p}\Big\{\left(\frac{1}{E_{p}}+\hbar\frac{1}{2mE^{(0)2}_{p}}\epsilon^{ij}\sigma_{i}p_{j}\right)\partial_{0}+\left[\frac{p^{i}}{E^{2}_{p}}+\hbar\frac{1}{2m^{2}E^{(0)}_{p}}\epsilon^{ij}\left(\sigma_{0}v_{i}-\sigma_{i}\right)\right]\partial_{i}\\
&~~~~~~~~~~~~~+\epsilon\,Q\left[\frac{\widetilde{E}_{j}}{E_{p}}+\epsilon_{ij}\frac{p^{i}}{E^{2}_{p}}B+\hbar\frac{m}{E^{(0)2}_{p}}b^{0}B\,\sigma_{j}+\hbar\frac{1}{2m^{2}E^{(0)2}_{p}}\sigma\cdot\widetilde{F}\,p_{j} \right]\partial^{j}_{p}  \Big\}f^{\epsilon}(x,\mathbf{p})
=0,
\end{split}\end{align}
where $v^{i}=-\partial^{i}_{p}E^{(0)}_{p}=p^{i}/E^{(0)}_{p}$ is the zeroth oder group velocity, $\widetilde{E}_{j}=E_{j}+\frac{1}{\epsilon\,Q}\partial_{j}E_{p}$ is the effective electric field. 
The effective energy is $E_{p}=E^{(0)}_{p}+\hbar\frac{\epsilon\,Q}{2m}\frac{p\cdot\widetilde{F}}{E^{(0)}_{p}}$. 

Taking the Taylor expansion to the $1/E_{p}$ and $1/E^{2}_{p}$ terms with keeping up to $\hbar$ order, one can rewrite the above equation as
\begin{align}\begin{split}\label{eq:ke}
&\Big\{\,\left(1+\hbar\epsilon^{ij}\sigma_{i}b_{j}\right)\partial_{0} \quad+\quad 
\frac{1}{\sqrt{G}}\left[\left(1-2\hbar\epsilon\,Q\,b\cdot \widetilde{F}\right)v^{j}+\hbar\frac{E^{(0)}_{p}}{m}\epsilon^{ij}\left(\sigma_{0}b_{i}-\sigma_{i}b_{0}\right)\right]\partial_{j}\\
&\;\;+\frac{\epsilon\,Q}{\sqrt{G}}\Big[\widetilde{E}_{j}+
B\epsilon_{ij}v^{i}-\hbar\epsilon\,Q(b\cdot \widetilde{F})(\widetilde{E}_{j}+2B\epsilon_{ij}v^{i})+\hbar\frac{1}{2E^{(0)2}_{p}}B\,\sigma_{j}+\hbar\frac{E^{(0)}_{p}}{m}(\sigma\cdot\widetilde{F})b_{j}\Big]\partial^{j}_{p}\, \Big\}f^{\epsilon}(x,\mathbf{p})=0
\end{split}\end{align}
Herein, for the sake of simplification, we have introduced a new vector $b^{\mu}=p^{\mu}/(2mE^{(0)2}_{p})$.
The energy $E_{p}$ and the factor $\sqrt{G}$ now can be written as, 
\begin{align}\label{eq:energy}
&E_{p}=E^{(0)}_{p}\left(1+\hbar\epsilon\,Qb\cdot\widetilde{F}\right) ,~~~~\sqrt{G}=1-\hbar\epsilon\,Qb\cdot \widetilde{F}.
\end{align}

Accordingly, the corresponding gap equation in Eq.~(\ref{eq:tq}) can also be reduced as the following by integration over $p_{0}$,
\begin{align}\begin{split}\label{eq:gapfull}
&m-m_{0}\\
&=-G\sum_{\epsilon=\pm1}\int \frac{\mathrm{d}^{2}p}{(2\pi)^22E^{(0)}_p}\bigg[2m\widetilde{f}^{\epsilon}(x,\mathbf{p})+\hbar\,\epsilon\,Q\frac{p\cdot\widetilde{F}}{E^{(0)}_{p}}\frac{d}{dE^{(0)}_{p}}\widetilde{f}^{(0)\epsilon}(x,\mathbf{p})-\hbar\,\epsilon\,Q\frac{B\,E^{(0)}_{p}+p\cdot\widetilde{F}}{E^{(0)2}_{p}}\widetilde{f}^{(0)\epsilon}(x,\mathbf{p})\bigg]\\
&~~~-\hbar\frac{GQB}{4\pi},
\end{split}\end{align}
where $\widetilde{f}^{+}(x,\mathbf{p})=f^{+}(x,\mathbf{p})$ and $\widetilde{f}^{-}(x,\mathbf{p})=f^{-}(x,\mathbf{p})-1$. The Eq.~(\ref{eq:ke}) and Eq.~(\ref{eq:gapfull}) are the  complete equal-time transport equation in 2+1 dimensions. 

\section{Transport equation with collision term }\label{sec03}

The quantum transport equation derived in the above section did not consider the collision term. To simply investigate the effect of the collision term, we will use the well-known relaxation time approximation. As the first step, the relaxation time approximation for Wigner function can be written as follows~\cite{Hakim1992,Yang:2003pz},
\begin{align}\label{eq:EOMWR}
\left(\slashed{K}-M\right)W(x,p)=-\frac{i\hbar}{2} \gamma\cdot\,u\frac{W(x,p)-W_{eq}}{\tau}\,,
\end{align}
where $u^{\mu}$ is the four fluid velocity of the hot medium, which can be determined by the Landau matching condition (such as $u\cdot J_{eq}=u\cdot J, \text{or } u_{\mu}T^{\mu\nu}_{eq}=u_{\mu}T^{\mu\nu}$), and $\tau$ is the relaxation time, which may depend on the space-time. There is a detailed analysis about the relaxation time approximation for Wigner function in~\cite{Hakim1992}. 

Then the kinetic equations for the 4 independent components now can be  written as
\begin{align}\begin{split}\label{eq:wcqr}
&\pi^{\mu}\mathscr{V}_{\mu}-M_{1}\mathscr{F}=0,\\
&\frac{1}{2}\hbar\triangledown^{\mu}\mathscr{V}_{\mu}+M_{2}\mathscr{F}=-\frac{\hbar}{2}u\cdot\frac{\mathscr{V}-\mathscr{V}_{eq}}{\tau},\\
&\pi_{\mu}\mathscr{F}-M_{1}\mathscr{V}_{\mu}+\frac{1}{2}\hbar\epsilon_{\mu\rho\sigma}\triangledown^{\rho}\mathscr{V}^{\sigma}=-\frac{\hbar}{2}\epsilon_{\mu\rho\sigma}u^{\rho}\frac{\mathscr{V}^{\sigma}-\mathscr{V}^{\sigma}_{eq}}{\tau},\\
&\frac{1}{2}\hbar\triangledown_{\mu}\mathscr{F}+M_{2}\mathscr{V}_{\mu}-\epsilon_{\mu\rho\sigma}\pi^{\rho}\mathscr{V}^{\sigma}=-\frac{\hbar}{2}u_{\mu}\frac{\mathscr{F}-\mathscr{F}_{eq}}{\tau}.
\end{split}\end{align} 

Similarly with the above section, using the semi-classical expansion method to solve this set of equations up to $\hbar$ order, we find the formal solution of scalar and vector component as follows
\begin{align}\begin{split}
\mathscr{F}&=m\,f(x,p)\delta(p^{2}-m^{2})-\hbar\,Qp\cdot \widetilde{F}\,f^{(0)}(x,p)\delta^{'}(p^{2}-m^{2}),\\
\mathscr{V}^{\mu}&=p^{\mu}f(x,p)\delta(p^{2}-m^{2})+\frac{\hbar}{2m}\epsilon^{\mu\rho\sigma}p_{\sigma}\left(\triangledown_{\rho} f^{(0)}(x,p)\right)\delta(p^{2}-m^{2})-\hbar\,mQ\widetilde{F}^{\mu}f^{(0)}(x,p)\delta^{'}(p^{2}-m^{2})\nonumber\\
&~~-\hbar\epsilon^{\mu\rho\sigma}p_{\sigma}\sigma_{\rho}\,f^{(0)}(x,p)\delta^{'}(p^{2}-m^2)+\frac{\hbar}{2m}\epsilon_{\mu\rho\sigma}u^{\rho}p^{\sigma}\frac{f^{(0)}-f^{(0)}_{eq}}{\tau}\delta(p^{2}-m^{2}).
\end{split}\end{align}
These equations are same with the solutions in Eq.(\ref{eq:F01}-\ref{eq:v0}) and Eq.(\ref{eq:solution-1}), respectively in zeroth order and $\hbar$ order, except the last term of the vector component $\mathscr{V}^{\mu}$. 

Furthermore, the covariant quantum transport equation can be derived as
\begin{align}\begin{split}\label{eq:tqr}
&\delta\left(p^{2}-m^{2}-\hbar\frac{Q}{m}p\cdot \widetilde{F}\right)\left[p\cdot\triangledown+m\sigma_{\nu}\partial^{\nu}_{p}+\hbar\frac{Q}{2m}\left(\partial_{\nu}p\cdot\widetilde{F}\right)\partial^{\nu}_{p}-\hbar\frac{1}{2m^{2}}\epsilon^{\mu\rho\sigma}\sigma_{\mu}p_{\sigma}\triangledown_{\rho}\right]f(x,p)\\
&=-\left(p\cdot\,u-\frac{\hbar}{2m^{2}}\epsilon_{\mu\rho\sigma}\sigma^{\mu}u^{\rho}p^{\sigma}+\hbar\frac{p\cdot\omega}{m}-\frac{\hbar}{2m}\epsilon_{\mu\rho\sigma}\frac{\partial^{\mu}\tau}{\tau}u^{\rho}p^{\sigma}  \right)\frac{f(x,p)-f_{eq}(x,p)}{\tau}\delta\left(p^{2}-m^{2}-\hbar\frac{Q}{m}p\cdot \widetilde{F}\right),
\end{split}\end{align}
where $\omega^{\mu}=(1/2)\epsilon^{\mu\rho\sigma}\partial_{\rho}u_{\sigma}$ is the vorticity vector. We find that the collision term in relaxation time approximation does not modify the on-shell condition. It is also obvious that this equation returns to the traditional relaxation time formalism of the kinetic equation when $\sigma^{\mu}, \omega^{\mu}, \partial^{\mu}\tau \rightarrow 0$. In addition, the corresponding gap equation is irrelevant to the relaxation time approximation by definition, hence it takes the same formula as that in the above section,
\begin{align}
	&m-m_{0}=-G\left[m\int \mathrm{d}^{3}p\,f(x,p)\delta(p^{2}-m^{2})-\hbar\,Q\int \mathrm{d}^{3}p\,p\cdot\widetilde{F}\,f^{(0)}(x,p)\delta^{'}(p^{2}-m^{2})\right].
\end{align}

Finally, the equal-time transport equation now can be written as 
\begin{align}\begin{split}\label{eq:kefr}
&\Big\{\,\left(1+\hbar\epsilon^{ij}\sigma_{i}b_{j}\right)\partial_{0}\\
&+\frac{1}{\sqrt{G}}\left[\widetilde{v}^{j}-\hbar\epsilon\,Qb_{0}\epsilon^{ij}\widetilde{E}_{i}+\hbar\epsilon\,Q\left(\epsilon^{ik}\widetilde{E}_{i}\,b_{k}\right)\widetilde{v}^{j}+2\hbar\epsilon\,QB\,b^{j}+\frac{1}{2E^{(0)2}_{p}}\left(\epsilon^{ik}\widetilde{v}_{i}\,\sigma_{k}\right)\widetilde{v}^{j}+\hbar\frac{E^{(0)}_{p}}{m}\epsilon^{ij}\left(\sigma_{0}b_{i}-\sigma_{i}b_{0}\right)\right]\,\partial_{j}\\
&+\frac{\epsilon\,Q}{\sqrt{G}}\Big[\widetilde{E}^{j}+
B\epsilon^{ij}\widetilde{v}_{i}+\hbar\epsilon\,Q\left(2b_{0}B+\epsilon^{kl}\widetilde{E}_{k}\,b_{l}\right)\left(\widetilde{E}^{j}+
B\epsilon^{ij}\widetilde{v}_{i}\right)+\hbar\frac{E^{(0)}_{p}}{m}(\sigma\cdot\widetilde{F})b_{j}\Big]\partial_{j}^{p} \Big\}\,f^{\epsilon}(x,\mathbf{p})\\
&=-\frac{1}{E_{p}}\left(p\cdot\,u-\frac{\hbar}{2m^{2}}\epsilon_{\mu\rho\sigma}\sigma^{\mu}u^{\rho}p^{\sigma}+\hbar\frac{p\cdot\omega}{m}-\frac{\hbar}{2m}\epsilon_{\mu\rho\sigma}\frac{\partial^{\mu}\tau}{\tau}u^{\rho}p^{\sigma}  \right)\frac{f(x,\mathbf{p})-f_{eq}(x,\mathbf{p})}{\tau}
\end{split}\end{align}
Similar, the corresponding gap equation remains the same:
\begin{align}\begin{split}\label{eq:gapfullr}
&m-m_{0}\\
&=-G\sum_{\epsilon=\pm1}\int \frac{\mathrm{d}^{2}p}{(2\pi)^22E^{(0)}_p}\big[2m\widetilde{f}^{\epsilon}(x,\mathbf{p})+\hbar\,\epsilon\,Q\frac{p\cdot\widetilde{F}}{E^{(0)}_{p}}\frac{d}{dE^{(0)}_{p}}\widetilde{f}^{(0)\epsilon}(x,\mathbf{p})-\hbar\,\epsilon\,Q\frac{B\,E^{(0)}_{p}+p\cdot\widetilde{F}}{E^{(0)2}_{p}}\widetilde{f}^{(0)\epsilon}(x,\mathbf{p})\big]\\
&~~~-\hbar\frac{GQB}{4\pi}.
\end{split}\end{align}
So far, we have developed the theoretical framework, as the combination of Eq.~(\ref{eq:kefr}) and Eq.~(\ref{eq:gapfullr}), to describe the evolution of distribution function $f$ for fermions with dynamical mass $m$ in 2+1~D.
The space-time evolution of such systems with any initial condition can be studied by solving the equation of motions numerically.

\section{The gap equation in equilibrium state}\label{sec04}

An interesting question is how the dynamical mass $m$ changes with temperature, chemical potential, and external field. In this section, we consider a simple case in which the system is under a constant electromagnetic field, and close to the global equilibrium state, as well as the condensation $\sigma$ is constant for space-time, i.e $\sigma_{\mu}=\partial_{\mu}\sigma(x)=0$. Besides, we set the mass $m_{0}=0$ for a clear physical picture. The transport equation Eq.~(\ref{eq:ke}) can be reduced to the following 
\begin{align}\begin{split}\label{eq:kes}
		&\Big\{\partial_{0}+\frac{1}{\sqrt{G}}\left(1-2\hbar\epsilon\,Q\,b\cdot\widetilde{F}\right)v^{i}\,\partial_{i}\\
		&~~~~~+\frac{\epsilon\,Q}{\sqrt{G}}\Big[\widetilde{E}_{j}+
		B\epsilon_{ij}v^{i}-\hbar\epsilon\,Q(b\cdot\widetilde{F})(\widetilde{E}_{j}+2B\epsilon_{ij}v^{i})\Big]\partial^{j}_{p} \Big\}f^{\epsilon}(x,\mathbf{p})=0.
\end{split}\end{align}
As discussed in appendix \ref{appendix-equilibrium function} , the equilibrium distribution function can be written as following, 
\begin{align}\begin{split}\label{eq:ef}
		&f^{\epsilon}(x,\mathbf{p})=\frac{1}{e^{(E_p-\epsilon\mu)/T}+1}=f^{(0)\epsilon}(x,\mathbf{p})+\hbar\frac{\epsilon\,Q}{2m}\frac{p\cdot \widetilde{F}}{E^{(0)}_p}\partial_{E^{(0)}_p}f^{(0)\epsilon}(x,\mathbf{p}),\\
		&f^{(0)\epsilon}(x,\mathbf{p})=\frac{1}{e^{(E^{(0)}_p-\epsilon\mu)/T}+1}.
\end{split}\end{align}
Herein, $f^{(0)\epsilon}(x,\mathbf{p})$ is the zeroth order equilibrium distribution function and $f^{\epsilon}(x,\mathbf{p})$ is the complete equilibrium distribution function which include the zeroth and first order contribution.
The notations $T$ and $\mu$ are the temperature and effective chemical potential, respectively. It should be noticed that we have let $u^{\mu}=(1,\mathbf{0})$ just for convenience, and the effective chemical potential $\mu(x)=\mu_{0}-QA_{0}$. Where $\mu_{0}$ is the chemical potential of fermions, and $A_{0}$ the electric potential. However, this conflicts with the assumption of constant $\sigma$. Therefore, the electric field should be absent and the effective chemical potential $\mu(x)=\mu_{0}$. Then, the gap equation Eq.~(\ref{eq:gapfull}) can be written as
\begin{align}\label{eq:gaps}
	&m=-G\sum_{\epsilon=\pm1}\int  \frac{\mathrm{d}^{2}p}{(2\pi)^22E^{(0)}_p}\left[2m\widetilde{f}^{\epsilon}(x,\mathbf{p})
	-\hbar\epsilon\,QB\frac{d}{dE^{(0)}_p}f^{(0)\epsilon}(x,\mathbf{p})\right]-\hbar\frac{GQB}{4\pi}.
\end{align} 
Herein, the the effective chemical potential $\mu(x)=\mu_{0}$ in the equilibrium distribution function. It means that there is no magnetic field effect on the zeroth order distribution function, which can be understood due to the system as a whole is under a static state because of the fluid velocity $u^{\mu}=(1,\mathbf{0})$. It may be noted that taking the zero temperature of the above result shows explicitly that the vacuum condensate is proportional to the magnetic field strength.

Substituting Eq.~(\ref{eq:ef}) into the gap equation Eq.~(\ref{eq:gaps}), one can find the analytical expression as
\begin{align}\label{eq:gapeq}
	&\frac{m^{2}}{2\pi}+m\left(\frac{1}{G}-\frac{\Lambda}{2\pi}\right)+\frac{m}{2\pi}T\sum_{\epsilon=\pm1}\ln\left(1+e^{(-m+\epsilon\mu)/T}\right)+\hbar\frac{QB}{2\pi}\left(\frac{1}{e^{(m-\mu)/T}+1}-\frac{1}{e^{(m+\mu)/T}+1} \right)+\hbar\frac{QB}{4\pi}=0.
\end{align}
It is obviously that the quantum correction is contributed by the magnetic field. The gap equation returns to the classical case when the magnetic field vanishes. 
The bare coupling constant $G$ can be fine-tuned, since the NJL model in 2+1 D is renormalizable. We take the normalization scheme as in Refs.~\cite{Rosenstein:1990nm,Cao:2014uva,Wang:2019nhd},
\begin{align}\label{eq:renomalization1}
	&\frac{1}{G}-\frac{1}{G_{c}}=-\frac{M_{0}}{2\pi}\mathrm{sgn}(G-G_{c}),
\end{align}
where the critical coupling $G_{c}=2\pi/\Lambda$ and $M_{0}>0$ are of finite quantities respectively, and $\mathrm{sgn}(x)$ is a sign function of $x$. For vacuum state in absence of the magnetic field, i.e. $QB=0$, $T = 0$, and $\mu = 0$, one can get 
\begin{align}\label{eq:gapvc}
	&\frac{m^{2}}{2\pi}+m\left(\frac{1}{G}-\frac{\Lambda}{2\pi}\right)=\frac{m^{2}}{2\pi}-\frac{m\,M_{0}}{2\pi}\mathrm{sgn}(G-G_{c})=0.
\end{align}
There are two solutions of the above equation, one is $m=0$, while another $m=M_{0}\,\mathrm{sgn}(G-G_{c})$. It means that the dynamical mass generation is only possible for  $G>G_{c}$, in which the dynamical mass $m=M_{0}$. The quantity $M_{0}$ plays a role as the effective fermion mass in vacuum. The solution $m=0$ is a trivial solution. This can be seen by introducing the thermodynamic potential $\Omega$. Path integral calculations~\cite{Miransky:2015ava,Cao:2014uva} show that $\partial{\Omega}/\partial\,m=\frac{m^{2}}{2\pi}-\frac{m\,M_{0}}{2\pi}\mathrm{sgn}(G-G_{c})$, hence Eq.(\ref{eq:gapvc}) is equivalent to the extremization condition of thermodynamic potential. For supercritical case ($G>G_c$), $\partial^{2}{\Omega}/\partial\,m^{2}|_{m=0}<0$ is the maximum value of the effective potential $\Omega$. While $\partial^{2}{\Omega}/\partial\,m^{2}|_{m=M_0}>0$ is the minimum of it. So $m=M_{0}$ is the physical mass. 
Besides, there is one, and only one trivial solution $m_0=0$ when the coupling constant $G$ equals to the critical coupling constant $G_{c}$. 

In the presence of the magnetic field, $QB \neq 0$, the situation is different and the dynamical mass generation can occur for arbitrary coupling constant $G$. In that case, the gap equation~(\ref{eq:gapeq}) becomes
\begin{align}\label{eq:gapvB}
	&\frac{m^{2}}{2\pi}-\frac{m\,M_{0}}{2\pi}\mathrm{sgn}(G-G_{c})+\hbar\frac{QB}{4\pi}=0\Rightarrow\eta^{2}_{m}-\eta_{m}\,\mathrm{sgn}(G-G_{c})+\frac{1}{2}\mathrm{sgn}(Q)\eta^{2}_{B}=0,
\end{align} 
where we defined the dimensionless variables $\eta_{m} \equiv m/M_{0}$ and $\eta_{B}\equiv \sqrt{|QB|}/M_{0}$. 
The corresponding solutions can be written as
\begin{align}
	&\eta_{m\pm}=\frac{1}{2}\left( \mathrm{sgn}(G-G_c)\pm\sqrt{\mathrm{sgn}(G-G_c)^{2}-2\mathrm{sgn}(Q)\eta^{2}_{B}} \right).
\end{align}
According to this equation, one can find that there is a non symmetry case for the sign of the electric charge $Q$. As mentioned before, this is due to the choice of irreducible representation of the Dirac matrices, which limits to the spin-up particles and spin-down anti-particles. Again, the physical solution of mass can be obtained by minimizing the effective potential $\Omega$. The left hand side of Eq.(\ref{eq:gapvB}) equals to $\partial{\Omega}/\partial\,m$. Firstly, in the case of negative charge ($Q<0$), the physical solution of mass m scaled by the vacuum mass $M_{0}$ are $\eta_{m+}=\frac{1}{2} (1+\sqrt{1+2\eta_{B}^{2}})$ for supercritical case ($G>G_{c}$), $\eta_{m+}=\frac{1}{2} (-1+\sqrt{1+2\eta_{B}^{2}})$ for subcritical case ($G>G_{c}$) and $\eta_{m+}=\frac{\eta_{B}}{2}$ for critical case ($G=G_{c}$). It means that the symmetry breaking can occurs for arbitrary magnetic field strength for these three different critical cases. Furthermore, the situation is different for the case of positive charge ($Q>0$), the physical solution are $\eta_{m+}=\frac{1}{2} (1+\sqrt{1-2\eta_{B}^{2}})$ for supercritical case ($G>G_{c}$) and $\eta_{m+}=\frac{1}{2} (-1+\sqrt{1-2\eta_{B}^{2}})$ for subcritical case ($G>G_{c}$). These shows us that the symmetry breaking can occurs for the case of the supercritical and subcritical in the case of positive charge, but they are limited to a small magnetic field strength, such as $0\le\eta_{B}\le\frac{1}{\sqrt{2}}$. However, there is no any solution for the case of critical case ($G=G_{c}$).


Now we move on to the more general case of finite temperature and chemical potential.
With the scheme of Eq.(\ref{eq:renomalization1}), the gap equation Eq.(\ref{eq:gapeq}) becomes 
\begin{align}\label{eq:gapeqrn}
	&\frac{m^{2}}{2\pi}-\frac{m\,M_{0}}{2\pi}\mathrm{sgn}(G-G_{c})+\frac{m}{2\pi}T\sum_{\epsilon=\pm1}\ln\left(1+e^{(-m+\epsilon\mu)/T}\right)+\hbar\frac{QB}{2\pi}\left(\frac{1}{e^{(m-\mu)/T}+1}-\frac{1}{e^{(m+\mu)/T}+1} \right)+\hbar\frac{QB}{4\pi}=0.
\end{align}
We can find that there are more rich phenomena of the symmetry breaking in the case of finite temperature. Similarly, he left hand side of Eq.(\ref{eq:gapeqrn}) equals to $\partial{\Omega}/\partial\,m$, and physical solutions of this equation minimize the effective potential $\Omega$.

\begin{figure}[!hbt]\centering
	\includegraphics[width=0.35\textwidth]{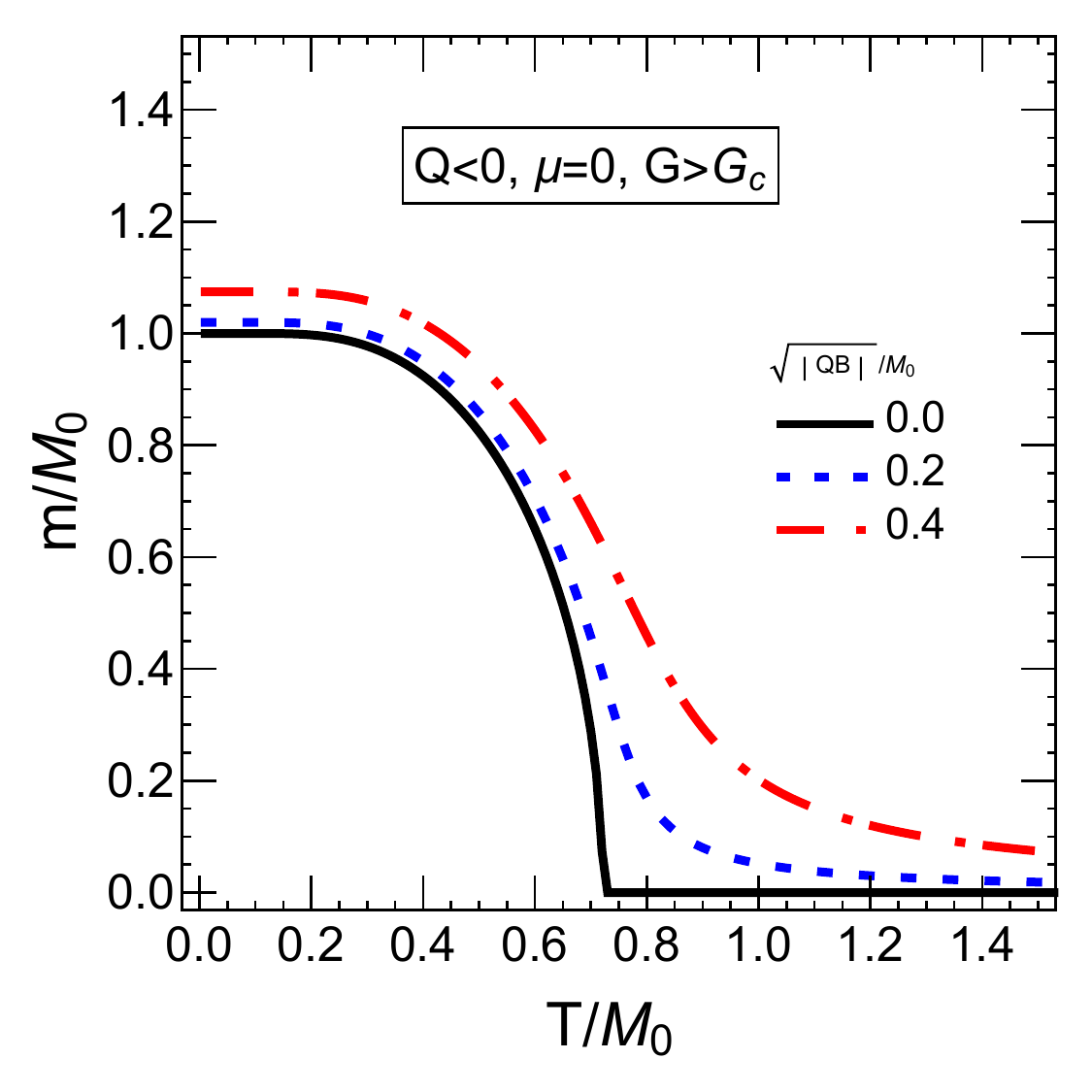}
	\includegraphics[width=0.35\textwidth]{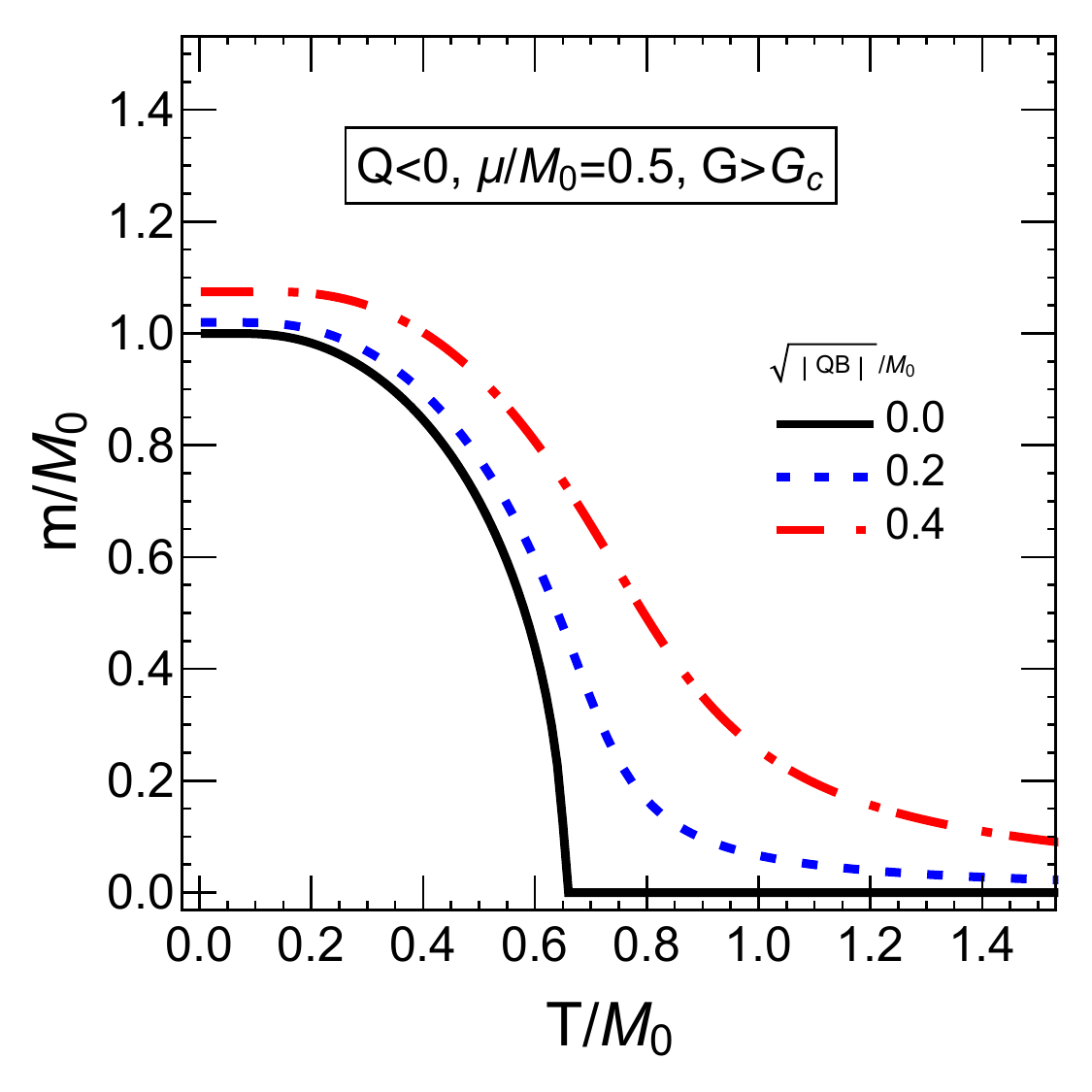}
	\caption{The dynamical mass $m$ as a function of the temperature $T$, scaled by the vacuum mass $M_{0}$.}
	\label{fig:gap1}
\end{figure}

\begin{figure}[!hbt]\centering
	\includegraphics[width=0.35\textwidth]{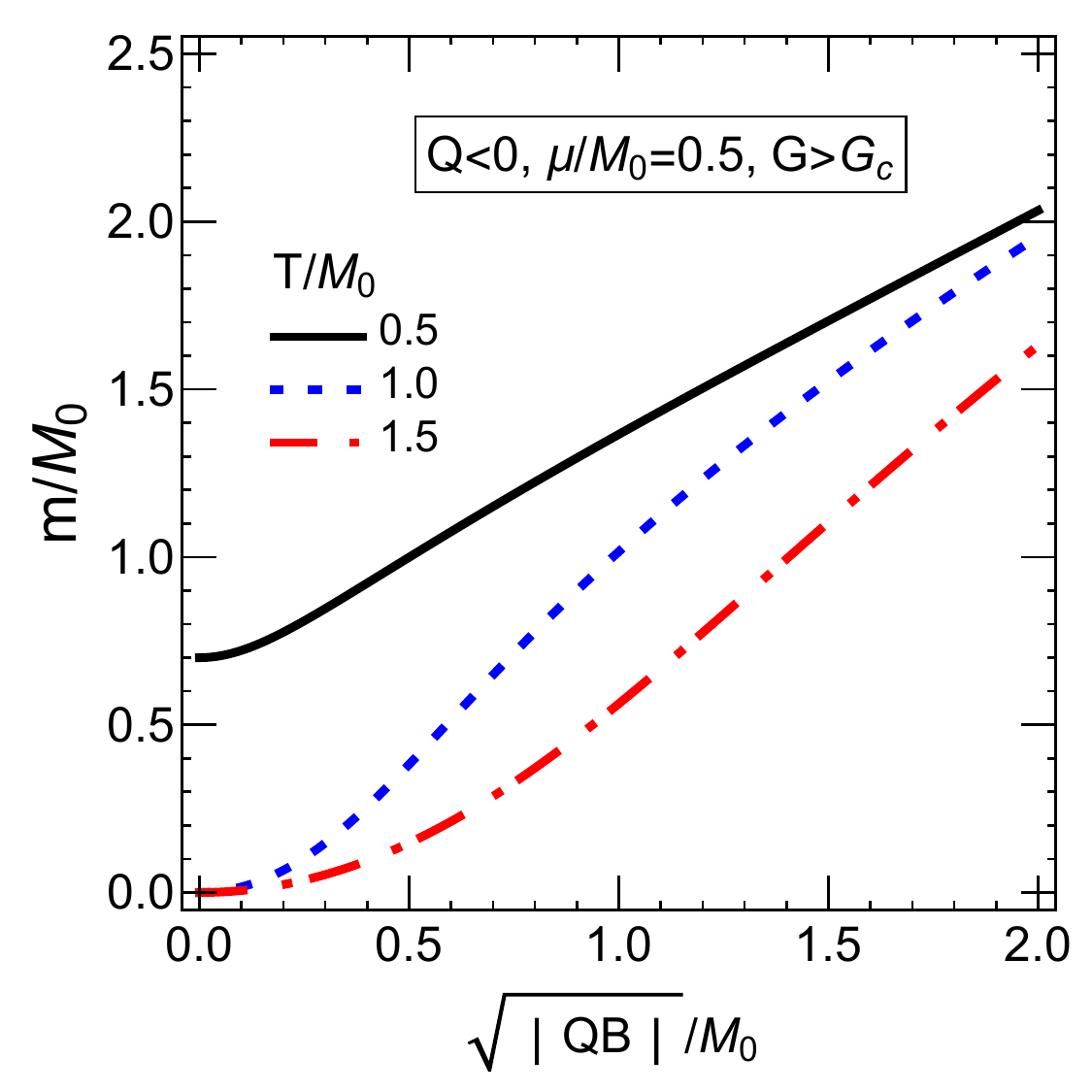}
	\caption{The dynamical mass $m$ as a function of the magnetic field $\sqrt{|QB|}$, scaled by the vacuum mass $M_{0}$. }
	\label{fig:gap3}
\end{figure}

\begin{figure}[!hbt]\centering
	\includegraphics[width=0.35\textwidth]{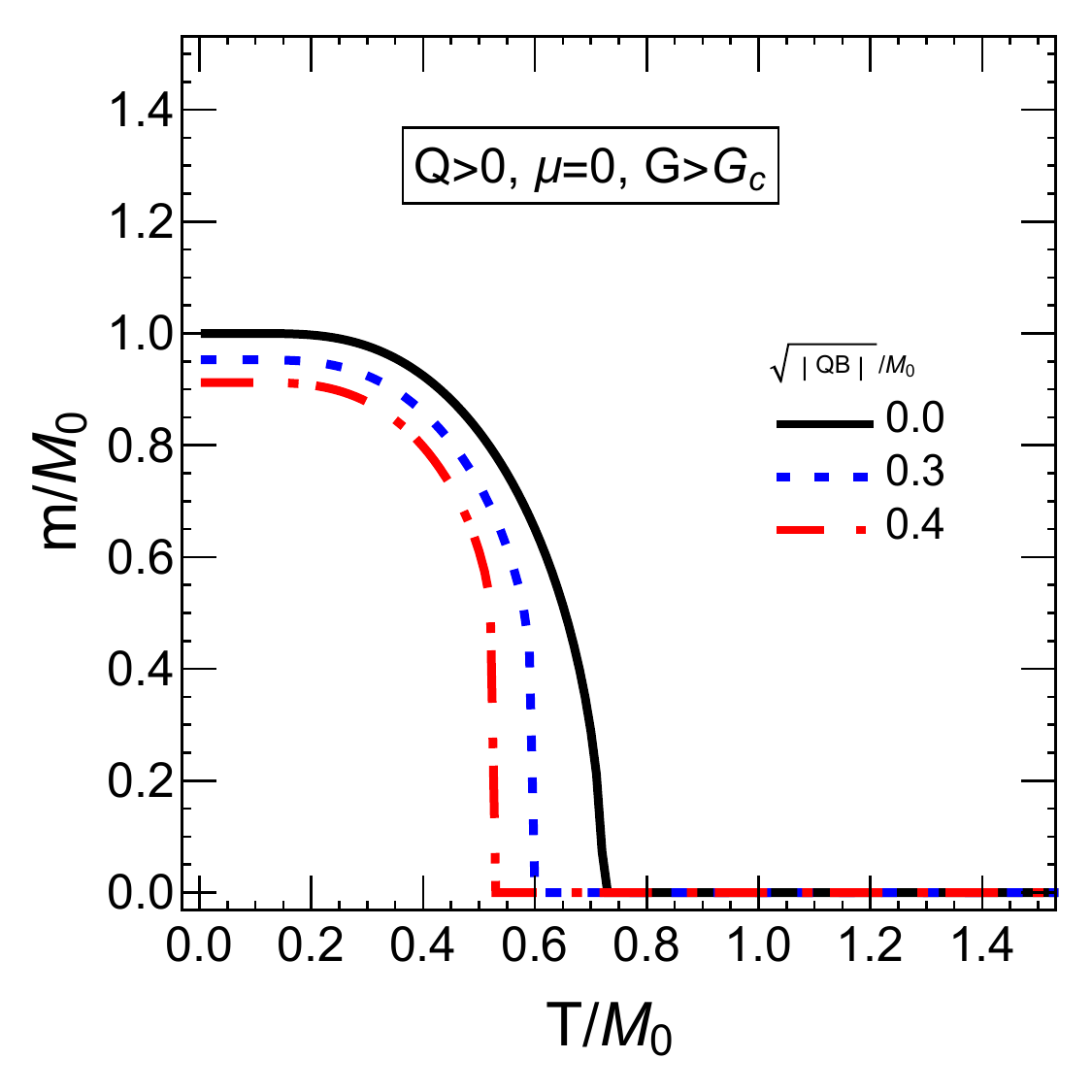}
	\caption{The dynamical mass $m$ is a function of the temperature $T$, scaled by the vacuum mass $M_{0}$ in case of $Q>0$.}
	\label{fig:gap4}
\end{figure}

The corresponding numerical results are showed in Fig.~\ref{fig:gap1}---\ref{fig:gap4}. Firstly, Fig.~\ref{fig:gap1} shows the dynamical mass $m$ is a function of the temperature, scaled by the vacuum mass $M_{0}$, for three different magnetic fields. The black solid line corresponds to the zero magnetic fields, which can be regarded as the classical results without the quantum correction. The dashed lines correspond to the results of $\sqrt{|QB|}/M_{0}=0.2$ and $0.4$, which includes the quantum correction originated from the interaction between the particle and magnetic field. While both of them correspond to $Q<0$ and $G>G_{c}$, the left panel is for neutral systems $\mu=0$ and the right panel is for a finite chemical potential $\mu/M_0=0.5$. We can find that dynamical mass is enhanced by the magnetic field and the finite chemical potential. We note that while the Lagrangian with $m_0=0$ has parity symmetry, the dynamical mass generation $m>0$ would break it spontaneously in the vacuum. Our results without magnetic field show that at high enough temperature the dynamical mass vanishes and the symmetry restores via a second order transition. Turning on a magnetic field, which explicitly breaks parity, causes the transition to become a crossover.


As mentioned previously, the dynamical mass generation is not identical for positive and negative charges, which can be seen by comparing Fig.~\ref{fig:gap4} and the right panel of Fig.~\ref{fig:gap1}. There is no non-trivial solution for the Eq.(\ref{eq:gapeqrn}) when the temperature $T$ beyond a given temperature, denoted by $T^{*}$, and the temperature $T^{*}$ is smaller for the stronger magnetic field. We can see that the dashed lines suddenly jump to zero beyond the temperature $T^{*}$, this is because that the left-hand side of Eq.(\ref{eq:gapeqrn}) is a monotonically increasing function beyond the temperature $T^{*}$, and the mass $m=0$ is corresponding to the minimum value of the thermal potential.

\section{Conclusion} \label{sec05}

In this work, we have derived the relativistic quantum kinetic equation for massive fermions with NJL interactions in 2+1~D from the Wigner function formalism by carrying out the semi-classical expansion up to $\hbar$ order. The equations are obtained both without and with a collision term. These results have allowed us to examine the quantum effect from electromagnetic fields on single-particle properties and to self-consistently obtain parity-odd transport currents induced by these external fields. By deriving the gap equation together with the transport equations, we have also investigated the dynamical mass generation phenomenon in this non-equilibrium framework. In particular, we've identified interesting quantum effects that are absent in the usual  classical mean-field result for the gap equation and that are  induced by the magnetic field and the collision term. We've  also found a new kind of quantum transport current that is induced by the gradient of out-of-equilibrium condensate. We've computed the mass gap in the special case of global equilibrium and constant magnetic field and found the nontrivial influence of the magnetic field on chiral condensate due to the quantum effects included in our results. 
As we've shown, the massive fermions in (2+1)D demonstrate interesting quantum features that are drastically different from the usual (3+1)D massive fermions. On the other hand, these features shown in the order-$\hbar$ transport equations also appear reminiscent of some properties seen in systems of (3+1)D massless chiral fermions. This may have its origin in the  correlation of fermion spin degree of freedom  with other degrees of freedom of the particles: in the (2+1)D massive case with particle/anti-particle due to dimensionality,  while in the (3+1)D massless case with momentum due to chirality. Such an interesting connection motivates further studies that we plan to carry out and report elsewhere in the future. We end this paper by noting again that a physical fermion state in (2+1)D could generally be    the superposition of two inequivalent irreducible representations as mirror images of each other. In the present work, we choose to  focus on one representation and understand the consequences of this specific mode alone. Studying   physical systems  with both sectors is certainly an interesting question which will be investigated as our next step. It would also be tempting to explore the possibility   that future developments may find certain (2+1)D quantum materials realizing an isolated sector.




\section*{Acknowledgments}
We thank Drs. Gaoqing Cao and Lingxiao Wang for helpful suggestions. 
AH, LH, XZ and PZ are supported by the NSFC Grants 11622539, 11775123, and 11890712. AH and JL acknowledges support by the NSF Grant No. PHY-1913729 and by the U.S. Department of Energy, Office of Science, Office of Nuclear Physics, within the framework of the Beam Energy Scan Theory (BEST) Topical Collaboration. 
SS is grateful to the Natural Sciences and Engineering Research Council of Canada. 

\begin{appendix}

\section{The definition of spin in the 2+1 D}\label{appendix-spin}
In the 2+1D systems, the spin operator is defined as $\sigma_z /2$, and 
there are two ways to show that such operator has the physical meaning of spin.
On one hand, $\sigma_z /2$ is the operator associated with rotational generator.
Under a Lorentz transformation, the spinor transforms as~\cite{Binegar:1981gv}  
\begin{align}
&\psi'(x)=S(\omega)\psi(x),\qquad
S(\omega)=e^{\frac{i}{2}\omega_{\mu\nu}J^{\mu\nu}},
\end{align}
where the $J^{\mu\nu}=\frac{i}{4}[\gamma^{\mu}, \gamma^{\nu}]$ is the generators of the Lorentz group (i.e $\mathbf{SO}(1,2)$). 
Noting that $J^{\mu\nu} = -J^{\nu\mu}$, there are three independent generators: 
\begin{align}
&
N^{1}=J^{10}=\frac{i\sigma_{2}}{2},\qquad
N^{2}=J^{20}=\frac{i\sigma_{1}}{2},\qquad
M=J^{12}=\frac{\sigma_{z}}{2}.
\end{align}  
While $N^1$ and $N^2$ are related to the boost transformation, $M\equiv\sigma_z /2$ is related to rotation.

On the other hand, $\sigma_z /2$ is related to the magnetic moment of the spinor, which can be seen from its coupling with the electromagnetic fields. Multiplying the Dirac equation under external fields by the operator $(i\slashed{D}+m)$, leads to the Klein-Gordon equation controlling the particle energy, 
\begin{align}
&(i\slashed{D}+m)(i\slashed{D}-m)\psi=\left[\left(iD\right)^2-\frac{Q}{2}\sigma^{\mu\nu}F_{\mu\nu}-m^2\right]\psi=\left[\left(iD\right)^2-Q\left(i\sigma^{1}E^{2}-i\sigma^{2}E^{1}-B\sigma_{z}\right)-m^2\right]\psi=0,
\end{align}
where $\sigma^{\mu\nu}=\frac{i}{2}[\gamma^{\mu},\gamma^{\nu}]$. So we can also define the spin operator as $M=\sigma^{12}/2=\sigma_{z}/2$ by the second term in the first identity and the magnetic field term in the second identity. 

Then, we move on to clarify the meaning of the statement ``the particle with spin up or the anti-particle with spin down''. 
Let us start from the free Dirac equation
\begin{align}
&(i\slashed{\partial}-m)\psi=0,
\end{align}
and work in one of the irreducible representations, named A, in which $\gamma^{0}_{A}=\sigma_{z}, \gamma^{i}_{A}=i\sigma^{i}$. 
In the momentum space, the positive and negative solutions to the Dirac equations are
\begin{align}
&\left(\slashed{p}-m\right)u(p)=0, \qquad
\left(\slashed{p}+m\right)v(p)=0.
\end{align}
In the particle rest frame, one can find the solutions respectively to be 
\begin{align}
&u(m,\mathbf{0})=\begin{pmatrix}1\\0\end{pmatrix},\qquad
v(m,\mathbf{0})=\begin{pmatrix}0\\1\end{pmatrix}.
\end{align}
It is straight-forward to see that they are eigenstates of spin operator,
\begin{align}
&M\,u(m,\mathbf{0})=+\frac{1}{2}u(m,\mathbf{0}),\qquad
M\,v(m,\mathbf{0})=-\frac{1}{2}v(m,\mathbf{0}).
\end{align}
Noting the $\pm{1\over2}$ eigenvalues for positive/negative solutions, respectively, one can see that the irreducible representation A 
represents ``the particle with spin up or the anti-particle with spin down". 
Similarly, one can perform the same analysis to other irreducible representation, B, and obtain the states with opposite eigenvalues of the spin operator.

\section{The equilibrium distribution function}\label{appendix-equilibrium function}

In this appendix, we discuss the form of the equilibrium distribution function.
The equilibrium distribution function was obtained in~\cite{DeGroot:1980dk,Chen:2013dca} as
\begin{align}\begin{split}\label{ap:solution1}
	&f^{\epsilon}_{eq}(x,\mathbf{p})=\frac{2}{(2\pi)^2}\frac{1}{e^{p\cdot \beta-\epsilon\alpha}+1},\\
	&p\cdot\beta=p^{0}\beta_{0}+p^{i}\beta_{i},~~\beta^{\mu}=u^{\mu}\beta,~~\alpha=\mu\beta,~~\beta=1/T,
\end{split}\end{align}
where $T$, $\mu$ and $u^{\mu}$ are the temperature, chemical potential and velocity of fluid, respectively. The energy $p^{0}=E^{(0)}_{p}==\sqrt{\mathbf{p}^{2}+m^{2}}$ at  the classical level, $p^{0}=E_{p}=E^{(0)}_{p}\left(1+\hbar\epsilon\,Qb\cdot\widetilde{F}\right)$ at the quantum level, see Eq.(\ref{eq:energy}). It is important to point out that the equilibrium distribution function must be a solution of the transport equation Eq.(\ref{eq:kes}). Let us see what the conditions are for the solution. It can be determined by the zeroth order of this transport equation, i.e
\begin{align}\begin{split}
&\Big\{\partial_{0}+v^{i}_{0}\,\partial_{i}+\epsilon\,Q\left(E_{j}+
B\epsilon_{ij}v^{i}_{0}\right)\partial^{j}_{p} \Big\}f^{\epsilon}(x,\mathbf{p})=0\,.
\end{split}\end{align}
Insertion of the above equilibrium distribution function into this equation yields
\begin{align}\label{ap:condition1}
	&p^{\mu}p^{\nu}\partial_{\mu}\beta_{\nu}-\epsilon\,p_{\mu}(\partial^{\mu}\alpha+QF^{\mu\nu}\beta_{\nu})=0.
\end{align} 
Herein, $\beta=1/T$, $\beta_{\mu}=u_{\mu}\beta$ and $\alpha=\mu\beta$. From this we can get the following conditions
\begin{align}\begin{split}\label{ap:condition2}
	&\partial_{\mu}\beta_{\nu}+\partial_{\nu}\beta_{\mu}=0,\\
	&\partial^{\mu}\alpha+QF^{\mu\nu}\beta_{\nu}=0.
\end{split}\end{align}
The first equation is the Killing's equation. In this work, we consider the case without rotation, $\partial_{\mu}\beta_{\nu}-\partial_{\nu}\beta_{\mu}=0$, hence the velocity and the temperature are all independent of $x$. Then the second equation of the above becomes
\begin{align}\label{app:1}
	&\partial^{\mu}\mu+QF^{\mu\nu}u_{\nu}=0.
\end{align} 
Taking the derivative with respect to space-time, we can get
\begin{align*}
	&\partial^{\mu}\partial^{\nu}\mu+Q\partial^{\mu}F^{\nu\sigma}u_{\sigma}=0,\\
	&\partial^{\nu}\partial^{\mu}\mu+Q\partial^{\nu}F^{\mu\sigma}u_{\sigma}=0
\end{align*}
Combining these two equations yields 
\begin{align*}
	&u_{\sigma}(\partial^{\mu}F^{\nu\sigma}-\partial^{\nu}F^{\mu\sigma})=0.
\end{align*}
Using the  Bianchi identity leads to 
\begin{align}\label{ap:condition3}
	&DF^{\mu\nu}=0,
\end{align}
where the derivative operator $D=u\cdot \partial$.
The above equation leads to $DA^{\mu}=\partial^{\mu}\phi$, $\phi$ is an arbitrary function, and we choose a gauge-fixing that $\phi=0$. Eq.~(\ref{ap:condition3}) means that the condition of equilibrium is the electromagnetic field $F^{\mu\nu}$ is constant in time in the rest frame of the fluid as determined by velocity $u^{\mu}$. As a consequence, (\ref{app:1}) can be simplified as
\begin{align}
	&\partial^{\mu}(\mu+QA\cdot u)=0.
\end{align}
The solution is $\mu(x)=\mu_{0}-QA\cdot u$, where $\mu_{0}$ is a constant. $\mu_0$ can be interpreted as the Gibbs function per particle, or the chemical potential, while $\mu(x)$ is the effective chemical potential containing electric potential~\cite{DeGroot:1980dk}.

According to the above conditions (\ref{ap:condition1},\ref{ap:condition2},\ref{ap:condition3}), we can now prove that the distribution function (\ref{ap:solution1}) also satisfy the transport equation Eq.(\ref{eq:kes}):
\begin{align*}
&\Big\{\partial_{0}+\frac{1}{\sqrt{G}}\left(1-2\hbar\epsilon\,Q\,b\cdot\widetilde{F}\right)v^{i}\,\partial_{i}\\
&~~~~~+\frac{\epsilon\,Q}{\sqrt{G}}\Big[\widetilde{E}_{j}+
B\epsilon_{ij}v^{i}-\hbar\epsilon\,Q(b\cdot\widetilde{F})(\widetilde{E}_{j}+2B\epsilon_{ij}v^{i})\Big]\partial^{j}_{p} \Big\}f^{\epsilon}(x,\mathbf{p})\\
&=\left(1-2\hbar\epsilon\,Q(b\cdot\widetilde{F})\right)\left[p^{\mu}p^{\nu}\partial_{\mu}\beta_{\nu}-\epsilon\,p_{\mu}(\partial^{\mu}\alpha+QF^{\mu\nu}\beta_{\nu})\right]f^{'\epsilon}_{eq}(x,\mathbf{p})\\
&=0.
\end{align*}
In this calculation, we have used the relations $D\widetilde{F}^{\mu}=1/2\epsilon^{\mu\rho\sigma}DF_{\rho\sigma}=0$, and $f^{'\epsilon}_{eq}(x,p)=df^{\epsilon}_{eq}(x,p)/d(p\cdot\beta)$. In semi-classical expansion, the distribution (\ref{ap:solution1}) can also be expanded as the following form.
\begin{align}\begin{split}
&f^{\epsilon}(x,\mathbf{p})=\frac{1}{e^{(E_p-\epsilon\mu)/T}+1}=f^{(0)\epsilon}(x,\mathbf{p})+\hbar\frac{\epsilon\,Q}{2m}\frac{p\cdot \widetilde{F}}{E^{(0)}_p}\partial_{E^{(0)}_p}f^{(0)\epsilon}(x,\mathbf{p}),\\
&f^{(0)\epsilon}(x,\mathbf{p})=\frac{1}{e^{(E^{(0)}_p-\epsilon\mu)/T}+1}.
\end{split}\end{align}
Noting that the velocity field is a global constant, we take the local rest frame of the whole system, $u^{\mu}=(1,\mathbf{0})$, and correspondingly the effective chemical potential $\mu(x)=\mu_{0}-QA_{0}$, and it will be cast into $\mu(x)=\mu_{0}$, when the electric field is absent.

\end{appendix}

\bibliography{reference}
\bibliographystyle{unsrt}

\end{document}